\documentclass[pra,twocolumn,showpacs]{revtex4-2}

\usepackage{amsmath}
\usepackage{latexsym}
\usepackage{amssymb}
\usepackage{graphicx}
\usepackage[colorlinks=true, citecolor=blue, urlcolor=blue]{hyperref}
\usepackage{float}
\usepackage{amsfonts}
\usepackage{textcomp}
\usepackage{mathpazo}
\usepackage{comment}
\usepackage{xr}

\usepackage{bbm}

\usepackage{xcolor}
\definecolor{myurlcolor}{rgb}{0,.5,.5}
\definecolor{mycitecolor}{rgb}{0,.6,0}
\definecolor{myrefcolor}{rgb}{2,0,0}
\usepackage{hyperref}
\hypersetup{colorlinks,
linkcolor=myrefcolor,
citecolor=mycitecolor,
urlcolor=myurlcolor}

\usepackage[draft]{fixme}
\usepackage{amsmath,bbm}
\usepackage{graphicx}
\usepackage{amsfonts}
\usepackage{amssymb}
\usepackage{amsmath, amssymb, amsthm,verbatim,graphicx,bbm}
\usepackage{mathrsfs}
\usepackage{color,xcolor,longtable}

\usepackage{xr}
\makeatletter
\newcommand*{\addFileDependency}[1]{
  \typeout{(#1)}
  \@addtofilelist{#1}
  \IfFileExists{#1}{}{\typeout{No file #1.}}
}
\makeatother

\newcommand*{\myexternaldocument}[1]{
    \externaldocument{#1}
    \addFileDependency{#1.tex}
    \addFileDependency{#1.aux}
}

\myexternaldocument{manuscript_PRL}

\newcommand{\beq}[0]{\begin{equation}}
\newcommand{\eeq}[0]{\end{equation}}

\newcommand{\one}{\leavevmode\hbox{\small1\normalsize\kern-.33em1}}

\def\be{\begin{equation}}
\def\ee{\end{equation}}
\def\ben{\begin{eqnarray}}
\def\een{\end{eqnarray}}
\def\eea{\end{array}}
\def\bea{\begin{array}}

\newcommand{\Tr}[1]{\mathrm{Tr}#1}
\newcommand{\bei}{\begin{itemize}}
\newcommand{\eei}{\end{itemize}}
\newcommand{\ket}[1]{|#1\rangle}

\newcommand{\I}{\mathbbm{1}}

\renewcommand{\emph}[1]{\textbf{#1}}

\makeatletter
\newtheorem*{rep@theorem}{\rep@title}
\newcommand{\newreptheorem}[2]{%
\newenvironment{rep#1}[1]{%
 \def\rep@title{#2 \ref{##1}}%
 \begin{rep@theorem}}%
 {\end{rep@theorem}}}
\makeatother

\theoremstyle{plain}
\newtheorem*{thm*}{Theorem}
\newreptheorem{thm}{Theorem}

\theoremstyle{definition}

\theoremstyle{remark}

\usepackage[T1]{fontenc}

\begin{document}

\title{Using quantum nonlocality for device-independent confirmation of relativistic effects}
\author{Shubhayan Sarkar}
\email{shubhayan.sarkar@ulb.be}
\affiliation{Laboratoire d’Information Quantique, Université libre de Bruxelles (ULB), Av. F. D. Roosevelt 50, 1050 Bruxelles, Belgium}

\begin{abstract}	
Synchronizing clocks to measure time is a fundamental process underpinning every practical communication task from GPS to parallel computation. However, as the current protocols are based on classical communication between the sender and receiver, they are prone to simple attacks that could cause a slight delay in the signal which would then cause a massive error in further operations. In this work, we first explain a simple attack that in principle can cause an arbitrary delay in the signal between sender and receiver. We then propose a way to overcome this problem by using a recently contrived idea of device-independent certification which utilises quantum nonlocality. Consequently, clocks can be synchronized in a highly secure way without trusting any devices in the setup. We then extend this proposal to observe relativistic time dilation in a device-independent manner.  
\end{abstract}


\maketitle

{\it{Introduction---}}
Quantum communication leverages the principles of quantum mechanics to transmit information securely. It relies on quantum properties, such as entanglement and superposition, to encode and exchange data. At the core of these protocols are correlations, or joint probability distributions, between space-like separated parties that defy explanation through classical systems. There are a host of such communication schemes such as quantum teleportation, quantum dense coding, quantum secure direct communication (QSDC) and quantum key distribution (QKD) to name a few \cite{commbook1,Nielsen_Chuang_2010,Gisin_2002}. Most of these protocols rely on trusting some aspects of the devices involved in the setup.

Quantum communication protocols that ensure security or functionality without needing to trust the internal workings of the devices used in the experiment are termed device-independent (DI) quantum communication tasks. The security of these protocols is guaranteed based solely on observed correlations that violate Bell inequalities \cite{Bell66, NonlocalityReview}. There are only a few quantum communication tasks that are DI, in particular, we only identified QKD \cite{crypto1,crypto2} and recently QSDC \cite{QSDC1,QSDC2} has been extended to the DI regime. Consequently, these communication protocols are the most secure as of the current literature. 

In this work, we consider another highly relevant practical task that everyone utilises on a daily basis, which is locating themselves using the global positioning system (GPS). In a simplified way, the GPS works by a set of satellites transmitting their time and location to the user who then utilises this data to compute its position and also to correct its clock. However, the GPS communicates via classical signals which makes it prone to simple attacks which might be extremely fatal. For instance, a GPS time error of about 1 nanosecond can cause a distance error of about 3 meters. This is particularly dangerous with the rise of autonomous vehicles where one has to be accurate within a few centimetres. 

Using an additional device that distributes entangled systems between the satellite and the user, we provide a quantum protocol to address the above problem. We utilise the concept of self-testing \cite{Yao, Supi__2020} to show the security of our protocol. Self-testing is the strongest DI scheme to certify the underlying quantum states and measurements based solely on the observed statistics, without needing to trust the internal workings of the devices. Again, self-testing ensures the correct functioning of the quantum devices by leveraging correlations that violate Bell inequalities. We name this setup as device-independent quantum GPS or DIQGPS. This protocol can then be used to synchronize the clocks of the satellite and the user securely and can detect if an intruder has access to the signals between the satellite and the user.

Synchronizing clocks are at the heart of the theory of relativity. Consequently, we then extend the idea of DIQGPS to consider the scenario where the satellite is in motion at a constant velocity. By introducing motion, we can analyze and test the impact of relativistic effects on the user’s clock compared to when the clock was stationary. This enables us to experimentally observe phenomena predicted by the theory of relativity, specifically time dilation, which describes how a moving clock ticks more slowly relative to a stationary one. Importantly, our approach provides the first method to test such relativistic effects without requiring trust in any of the devices used in the experiment. This ensures that the observed results are solely based on fundamental physical principles and are free from any assumptions about the reliability or calibration of the equipment.

{\it{Synchronizing clocks---}} Before proceeding, let us clarify that synchronizing clocks in this work implies synchronizing the time interval of some event among spatially separated parties and not the absolute time of the clocks. For simplicity, we will assume that the absolute time of the clocks is the same at the start of the protocol. A simple way to do this is to bring the two clocks to the same location and then put the same time on both clocks. Moreover, we assume in this section that the clocks are rest with respect to each other. Here we focus on whether the time interval of some event measured by these clocks can be manipulated by some attacker which will also cause an error in recording their position with respect to each other.  For instance, if the time of two clocks are denoted by $t_1,t_2$, then if any of the two clocks record a time interval of some event as $\Delta t_i (i=1,2)$ in their frame of reference, then we want to ensure $\Delta t_1=\Delta t_2$ in a device-independent way, that is, without trusting or knowing the internal workings of any device involved in the experiment.

\begin{figure}
    \centering
    \includegraphics[width=.8\linewidth]{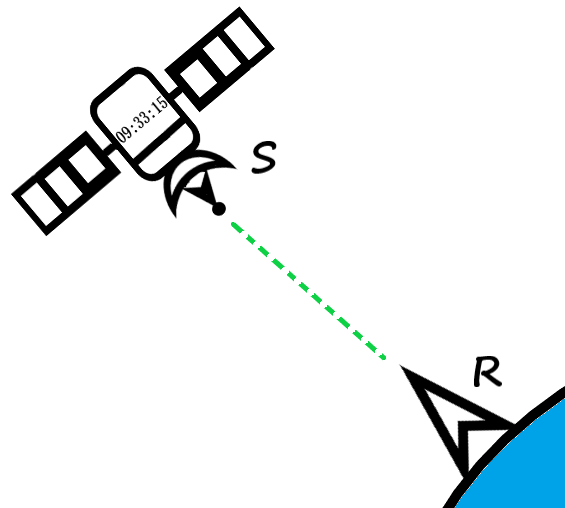}
    \caption{A simplified GPS setup involves a satellite $S$ transmitting the time of signal dispatch and a receiver $R$ noting the time of reception. Using the time difference and the speed of light, $R$ calculates its distance from $S$.}
    \label{fig1}
\end{figure}

Let us now begin by considering a simplified version of a GPS consisting of a single satellite $S$ and a receiver $R$. The satellite $S$ transmits a signal containing the precise time it was sent, according to $S'$s onboard atomic clock. The receiver $R$ records the time it received the signal using its own clock. Now, $R$ computes its distance to $S$ by calculating the signal's travel time which can be done by taking the difference between the transmission and reception times along with knowledge of the speed of light [see Fig. \ref{fig1}]. 

\begin{figure}
    \centering
    \includegraphics[width=.8\linewidth]{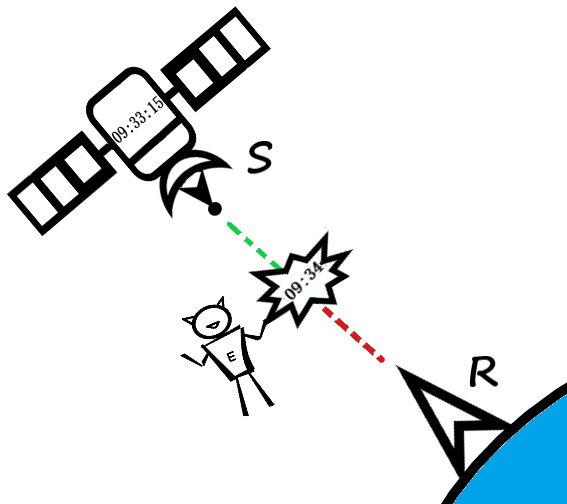}
    \caption{Signal injection or time manipulation attack on the GPS signal: The time sent by the GPS is manipulated by the attacker Eve by delaying the signal or changing the time stamp.}
    \label{fig2}
\end{figure}

Consider now a simple attack on the above setup. For this purpose, an eavesdropper, $E$, intercepts the signal from the satellite and then copies it. Then after a time lag, it sends a fake signal to $R$ with the same information as received by $S$ [see Fig. \ref{fig2}]. This kind of attack on a GPS is also known as signal injection. For a review on attacks on GPS refer to \cite{GPS1}. Consequently, even if $R$ and $S$  try to coordinate their data by some other method, the presence of the eavesdropper can never be detected. This would lead to the wrong reception time at $R$ and thus any further computation using this data will lead to incorrect results. This kind of attack is also possible if $R,S$ sets up a key to encrypt the time data. Another possible attack is that the eavesdropper provides completely fake data with the wrong time to the receiver. This type of attack is known as time manipulation. Let us now provide a protocol based on Bell's scenario \cite{Bell} that is secure against these types of attacks. Additionally, one does not even need to trust any devices involved in the setup to guarantee safety. 

\begin{figure}
    \centering
    \includegraphics[width=.8\linewidth]{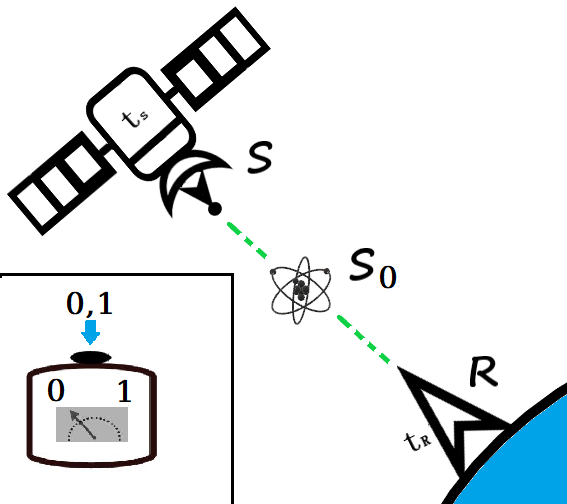}
    \caption{DIQGPS: $S_0$
  generates an entangled state sending one subsystem to the receiver $R$ and the other to satellite $S$. Both $R$ and $S$ perform two binary outcome measurements, with inputs $x,y=0,1$ and outcomes $r,s=0,1$, respectively (see bottom left). Additionally, $S$ and $R$ use clocks to record the times of their first two detections. $S$ encodes these times in the inputs of its subsequent measurements. Once the experiment is done, it sends all its data to $R$ and also reveals the rounds which were used to encode its time.}
    \label{fig3}
\end{figure}

{\it{Device-independent way to synchronize clocks---}}
 Let us modify the above-described scenario by considering two satellites $S,S_0$. Furthermore, the satellite $S_0$ consists of a source that can generate entangled state $\rho_{RS}$ and sends one of the subsystems to the receiver $R$ and the other to $S$. On their respective subsystems, both $R$ and $S$ perform two binary outcome measurements each denoted by the inputs $x,y=0,1$ with the outcomes denoted as $r,s=0,1$ respectively. Along with this, $S,R$ also have clocks that measure the time of respective first two detections denoted as $t_R^i,t_S^i (i=1,2)$ [see Fig. \ref{fig3}]. This will be referred to as the test phase of the scheme. It is important to recall here that $R$ and $S$ can not communicate with each other during this phase of the experiment. 

 The experiment is repeated enough times and after all the detection events, $S$ sends all its data to $R$ who then 
 constructs the joint probability distribution or correlations, $\vec{p}=\{p(r,s|x,y)\}$ where $p(r,s|x,y)$ denotes the probability of obtaining outcome $r,s$ by $R$ and $S$ when they choose the inputs $x,y$ respectively. Additionally, $S$ encodes the time of the first two detection events $t_S^i$ in the later input choices. For instance, if $S's$ clock records time $t=1,2$ for the first two detections, then it encodes this information by choosing the input in the third, fourth round as $0,1$ and fifth, sixth round as $1,0$ and then proceeds in a random manner (the chosen rounds is in general random). Once all the data is sent to $R$ then $S$ reveals the rounds which encoded the time of the detection events. This will be referred to as the reveal phase of the scheme. 
 We will call the above setup as device-independent quantum GPS or DIQGPS.
 
 The probabilities $p(r,s|x,y)$ can be computed in quantum theory using the Born rule as
\begin{eqnarray}
p(r,s|x,y)=\Tr\left[(N^{R}_{r|x}\otimes N^{S}_{s|y})\rho_{RS}\right]
\end{eqnarray}
where $N^{R}_{r|x},N^{S}_{s|y}$ denote the measurement elements corresponding to inputs $x,y$. The measurement elements are positive semi-definite and $\sum_{r}N^{R}_{r|x}=\sum_{s}N^{S}_{s|y}=1$ for all $x,y$.
It is usually simpler to express the probabilities in terms of expectation values as
\begin{eqnarray}\label{exp1}
\langle\mathcal{R}_{x}\mathcal{S}_{y}\rangle=\sum_{r,s=0,1}(-1)^{r+s}p(r,s|x,y)
\end{eqnarray}
where $\mathcal{R}_{x}, \mathcal{S}_{y}$ denote Alice's and Bob's observable corresponding to the input $x,y$ respectively. 

Now, using the correlations $\vec{p}$, $R$ computes the following functional:
\begin{eqnarray}
\mathcal{B}=\langle\mathcal{R}_{1}\mathcal{S}_{1}+\mathcal{R}_{1}\mathcal{S}_{2}+\mathcal{R}_{2}\mathcal{S}_{1}-\mathcal{R}_{2}\mathcal{S}_{2}\rangle.
\end{eqnarray}
This is the well-known Clauser-Horne-Shimony-Holt (CHSH)-Bell functional \cite{CHSH}. The maximum value of the above functional $\mathcal{B}$ that is attainable using classical strategies is $2$. However using the maximally entangled state $\ket{\psi_{RS}}=\ket{\phi^+_{RS}}=1/\sqrt{2}(\ket{00}+\ket{11})$ and the observables $\mathcal{\tilde{R}}_{1}=\sigma_z,\mathcal{\tilde{R}}_{1}=\sigma_x$ and $\mathcal{\tilde{S}}_{1}=(\sigma_z+\sigma_x)/\sqrt{2},\mathcal{\tilde{S}}_{2}=(\sigma_z-\sigma_x)/\sqrt{2}$ where $\sigma_z,\sigma_x$ are the Pauli observables, one can attain a value $2\sqrt{2}$.

Let us now recall device-independent certification, particularly the task of self-testing. For this purpose, we consider a quantum device performing a Bell experiment on some state $\ket{\psi_{RS}}\in\mathcal{H}_{R}\otimes\mathcal{H}_{S}$ using binary outcome observables $\mathcal{R}_x$ and $\mathcal{S}_y$, where the dimensions of $\mathcal{H}_{R}$ and $\mathcal{H}_{S}$ are unknown but finite. As the dimension is unrestricted, without loss of generality we consider that the state shared among $S$ and $R$ is pure. The only information one obtains from the devices is the correlation $\vec{p}$. Self-testing aims to reveal the form of the state and observables from $\vec{p}$. However, there are two degrees of freedom which can never be detected from the observed statistics: (i) local unitaries $U_{R},U_{S}$ that act on $\mathcal{H}_{R},\mathcal{H}_{S}$, that is, the state $U_{R}\otimes U_{S}|\psi_{RS}\rangle$ together with $\{U_{R}\mathcal{R}_xU^\dagger_{R}\},\{U_{S}\mathcal{S}_yU^\dagger_{S}\}$ will generate the same $\vec{p}$ (ii) the presence of an auxiliary system on which the measurements act trivially. There are numerous self-testing schemes (see, e.g., Refs. \cite{Scarani,Reichardt_nature,Mckague_2014,Wu_2014,Bamps,All,chainedBell, Projection, Armin1,sarkar,Allst,sarkar2024universal}) that provide self-testing statements for various states and measurements. In particular, Refs. \cite{Projection} and \cite{Allst,sarkar2024universalschemeselftestquantum} propose self-testing strategies for any bipartite or multipartite pure entangled states, where the second method is based on the quantum networks scenario.

If the value of $\mathcal{B}=2\sqrt{2}$, then one can infact certify that the state shared among $R,S$ is 
\begin{eqnarray}
    U_{R}\otimes U_{S}\ket{\psi_{RS}}=\ket{\phi^+_{R'S'}}\otimes\ket{\xi}_{R''S''}
\end{eqnarray}
where  $\mathcal{H}_S=\mathcal{H}_{S'}\otimes\mathcal{H}_{S''}, \mathcal{H}_R=\mathcal{H}_{R'}\otimes\mathcal{H}_{R''}$ such that $\mathcal{H}_{S'}=\mathcal{H}_{R'}=\mathbb{C}^2$ and $\ket{\xi}_{R''S''}$ is the auxiliary state. The observables are certified for all $x,y$ as
\begin{eqnarray}
    U_{R}\mathcal{R}_xU^\dagger_{R}=\mathcal{\tilde{R}}_x\otimes\I_{R''},\quad U_{S}\mathcal{S}_yU^\dagger_{S}=\mathcal{\tilde{S}}_y\otimes\I_{S''}.
\end{eqnarray}
For the proof of the above fact, refer to \cite{Scarani,PHDSARKAR}.
Using this fact, let us analyse how observing the Bell value $\mathcal{B}=2\sqrt{2}$ in the DIQGPS scheme allows us to synchronize clocks and then find the relative distance in a device-independent manner.

As described earlier, two major attacks can be done on the traditional GPS. Let us argue here that both of these attacks can be detected in the DIQGPS. First, due to the no-cloning theorem \cite{Wootters1982} in quantum theory, the attacker can not copy the entangled bits sent by $S_0$. Consequently, the attacker can not copy the qubits sent by $S_0$ and send them at a later time. As a matter of fact, if the attacker tries to interact with any of the entangled bits, then the quantum state would change and one can not observe the maximal violation of the CHSH-Bell inequality. Consequently, observing the value of $\mathcal{B}=2\sqrt{2}$ ensures that there can not be any intruder during the test phase of the experiment and that the times recorded by $R,S$ are accurate. Now, during the reveal phase since the time recorded by $S$ is communicated via its inputs which is unknown to the attacker. Consequently, there is no way for the attacker to change this information in a systematic manner but can only change the entire sequence of inputs. However, this will again lead to wrong $\vec{p}$ being computed by $R$ and thus the test would fail and the attacker would be detected. Consequently, neither signal injection nor time manipulation attack is possible in the DIQGPS and $S$ can safely communicate the time of the detection events to $R$.

After receiving the time of the first two detections of $S$ by $R$, it checks whether $t_R^2-t_R^1=t_S^2-t_S^1$. If not then $R$ adjusts its clock accordingly and thus their clocks will be synchronized. Furthermore, to compute the distance between $R,S$ it uses the fact that entangled photons are generated at the same point in space-time. For simplicity, we assume that all $S_0,S,R$ lie on the same line [see Fig. \ref{fig3}]. Let us say that the first entangled photons were ejected at $S_0$ at time $t_{S_0}$. Then, the distance between $R,S$ can be computed using
\begin{eqnarray}
    z_R=z_{S_0}+c(t_{R}^1-t_{S_0}),\quad z_S=z_{S_0}-c(t_{S}^1-t_{S_0})
\end{eqnarray}
where $z_{S_0},z_R,z_S$ are the position of $S_0,R,S$ and $c$ is the speed of light. Now, the distance between $R,S$ is simply given by
\begin{eqnarray}\label{eq7}
    z_R-z_S=c(t_{R}^1+t_{S}^1-2t_{S_0}).
\end{eqnarray} 
Consequently, the time of ejection of the entangled photons also has to be known by $R$ to compute its distance with $S$. A possible way to achieve this would be that the source $S_0$ is inside $S$ and thus using the time of arrival of the photon and knowing the distance between $S_0$ and its measurement device, $S$ can know the time $t_{S_0}$ and again encode it in its input choices as done before. Consequently, space-based entanglement distribution setups can be utilised for this purpose (e.g. Ref. \cite{spaceent1}). However, one has to ensure that along with the quantum bits, $S$ does not send its input choices during the test phase, as this would violate the measurement independence assumption in the Bell experiment and then classical systems could also reproduce the maximal quantum value. Another way would be that $S$ is in between $S_0$ and $R$ and thus the factor $t_{S_0}$ will be cancelled from \eqref{eq7}. 

{\it{Device-independent way to test time dilation---}}
\begin{figure}
    \centering
    \includegraphics[width=.8\linewidth]{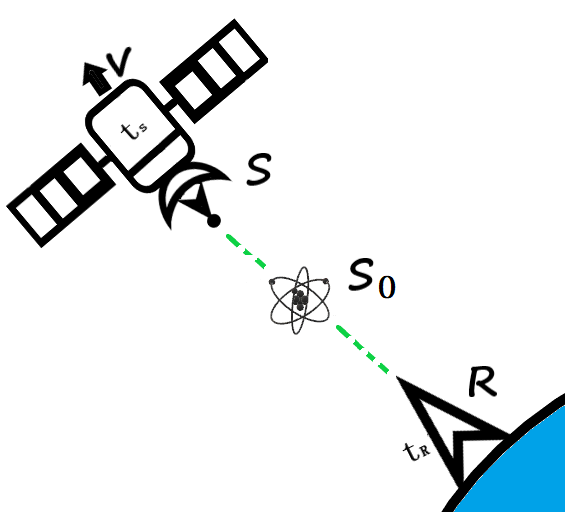}
    \caption{DIQGPS with $S$ moving at a constant velocity.}
    \label{fig4}
\end{figure}
Let us again consider the setup of DIQGPS but now $S$ starts to move with a constant speed of $v$ along the line as shown in Fig. \ref{fig4}. Now, we want to observe the time dilation of $S's$ clock with respect to $R's$ clock. In the first phase of the experiment, $S$ stays at rest and synchronizes its clock with $R$ using the above DIQGPS scheme. Then $S$ starts to move at a constant speed $v$ and the same procedure as DIQGPS is repeated. Now, considering the relativistic time dilation due to the motion of $S$, if the time between two signals is recorded as $t_R^2-t_R^1$ by $R$, then the detection of the corresponding entangled partners should happen at $S$ as
\begin{eqnarray}\label{eq8}
  t_S^2-t_S^1=  (t_R^2-t_R^1)\frac{c}{(c-v)\sqrt{1-v^2/c^2}}.
\end{eqnarray}
The computation of the above quantity in the scenario concerned is presented in Appendix A. Without relativistic effect, the factor $\sqrt{1-v^2/c^2}$ will be absent from the denominator of \eqref{eq8}. It is important to note here that we assume that the correlations $\vec{p}$ remain invariant when $S$ moves at a constant velocity. 
This is possible if the above setup is based on the polarization of photons, that is, the maximally entangled state is encoded via the horizontal and vertical polarisation of photons as done in the significant loophole-free experiment on observing Bell violation \cite{Giustina_2015}. This is because polarization remains invariant under a Lorentz boost in the same direction as the propagation of photons. However, this is not true for massive particles as spin and momentum are not independent of each other. 
Consequently, the above setup also opens up an interesting avenue for testing the effect of constant velocity on correlations generated by massive particles in a device-independent manner.

{\it{Discussions---}}
This work addresses vulnerabilities in GPS for positioning and clock synchronization, which relies on classical signals susceptible to attacks. To mitigate this, we proposed a device-independent quantum GPS (DIQGPS) that uses entangled systems and the technique of self-testing that certifies quantum states and measurements without trusting device internals, leveraging Bell inequality violations for security. DIQGPS enables secure clock synchronization and intrusion detection between satellites and users. 
The approach is further extended to test relativistic effects, such as time dilation, when the satellite is in motion. This novel method experimentally observes relativistic phenomena without relying on device trust, ensuring results are grounded in fundamental physics.

In a realistic DIQGPS setup, one would require at least three satellites to pin down the three-dimensional position of $R$. It can then be done by considering the same setup but the source now distributes a four-party maximally entangled state and then one can utilise either the self-testing of state using Mermin's inequality \cite{Jed3} or the one proposed in \cite{sarkar4}. One can alternatively utilize the same setup as the one proposed above but now three entangled sources will be required and $R$ has to measure three times more statistics which will be more resource-intensive. We also realize that this method will be much slower when compared to standard GPS as one has to gather a large amount of statistics and thus at the current stage from a practical perspective it might not be useful for real-time synchronization but will be useful in areas which require the highest security, for instance, in the military or aviation. Moreover, further work is required from a theoretical perspective to make the scheme robust to experimental errors and finite statistics and understand the impact of not observing the exact Bell value but a value closer to it. In the proposed setup one only has to record the time of arrival of the first two photons. As one attains the exact maximal quantum violation, this is fine as there are no losses or dark count events. However, in the practical scenario, one also has to record the time of every detection. In this work, we analysed when one of the parties moves at a constant velocity with respect to the other. However, in a practical setup, one must also analyse when all the parties are moving in a non-inertial frame of reference and consider the effects of transverse motion. 

\textit{Some implicit assumptions in the above scenario---}
\begin{itemize}
    \item (Assumption 1:) Eve has no quantum memory; thus, she can not store the quantum bit sent by the source and release it at a later time.
    \item (Assumption 2:) Authenticated classical channel is required (like quantum cryptography) to ensure that the commnication is genuinely between a satellite and the user, and not between an adversary and the user. This stops impersonation attacks by the adversary.
\end{itemize}

Given these two assumptions, one can ensure that there are no time-delay attacks in the GPS using the above scenario. Both of these assumptions might be well-justified in some particular scenario, but the scheme is not device-independent unless assumption 1 can be removed.

{\it{Acknowledgements---}}
 This project was funded within the QuantERA II Programme (VERIqTAS project) that has received funding from the European Union’s Horizon 2020 research and innovation programme under Grant Agreement No 101017733.

\providecommand{\noopsort}[1]{}\providecommand{\singleletter}[1]{#1}

\onecolumngrid
\appendix
\section{Time relation between $S$ and $R$ when $S$ moves at a constant speed $v$}
Let us suppose that the distance between $S$ and $S_0$ when the first entangled photons are ejected is $x_S$ and the time at the clocks at $R$ is denoted as $t_R^1,t_R^2$ for the first two photons detections. Let us consider the reference frame of $R$. 
$S$ moves at a constant speed $v$ with respect to $R$ and thus according to $R$ the time at taken by the first photon to reach $S$ denoted as $t_{S|R}$ is calculated as
\begin{eqnarray}\label{A1}
    x_S+v t_{S|R}^1=c t_{S|R}^1\implies x_S=t_{S|R}^1(c-v)
\end{eqnarray}
The second photon's arrival time at $R$ is $t_R^2$; thus, the time difference between the photons is $t_R^2-t_R^1$. Consequently, the time taken by the second photon to reach $S$ according to $R$ is given by
\begin{eqnarray}\label{A2}
    x_S+v(t_R^2-t_R^1)+v t_{S|R}^2=ct_{S|R}^2
\end{eqnarray}
Substituting $x_S$ from \eqref{A1} in \eqref{A2}, we obtain that
\begin{eqnarray}
   (c-v)(t_{S|R}^2-t_{S|R}^1)= v(t_R^2-t_R^1)\implies t_{S|R}^2-t_{S|R}^1=(t_R^2-t_R^1)\frac{v}{c-v}.
\end{eqnarray}
Adding now the time difference in the ejection times of the photons $t_R^2-t_R^1$, we obtain that time difference between the detection events at $S$ according to $R$ is $(t_R^2-t_R^1)\frac{c}{c-v}$.
Now, considering the relativistic effect on $S's$ clock, the time difference between the photons with respect to $S$ is given by
\begin{eqnarray}
    t_S^2-t_S^1=(t_R^2-t_R^1)\frac{c}{(c-v)\sqrt{1-v^2/c^2}}
\end{eqnarray}
which gives us the desired formula \eqref{eq8}.

\end{document}